\documentclass[twocolumn,showpacs,pre,aps]{revtex4}

\usepackage{dcolumn}
\usepackage{bm}

\begin{document}

\title{Stochastic Representation of Deterministic Interactions \\
and Brownian Motion}

\author{Yuriy E. Kuzovlev}
\email{kuzovlev@kinetic.ac.donetsk.ua}
\affiliation{A.A.Galkin Physics and Technology Institute
of NASU, 83114 Donetsk, Ukraine}

\date{\today}

\begin{abstract}
Exact generalized stochastic representation of deterministic
interaction between two dynamical (quantum or classical) systems
is derived which helps when considering one of them to replace
another by equivalent commutative ($c$-number valued) random
sources. The method is applied to classical Brownian motion of a
particle in a gas, and statistics of this motion is reduced to
statistics of the gas response to perturbations.
\end{abstract}


\maketitle

\section{Introduction}

In recent years the old variety of problems about interaction
between one or another ``dynamical system'' (DS) and its
environment (named also ``thermal bath'' or ``thermostat'') have
attracted new attention as a part of the quantum computation
problem. Frequently, the environment contains too many routine
degrees of freedom to be of special interest in itself, and one
would be satisfied by a simplified statistical description of DS
in terms of its own degrees of freedom only but with adding of
random sources, or ``Langevin forces'', which effectively
substitute the ``thermostat''. In this respect, during last
century many different approaches were proposed. Perhaps, the two
extreme points among them are the highly time-nonlocal description
of the DS evolution by means of path integrals and influence
functionals \cite{f}, on one hand, and the mathematical theory of
quantum semigroup evolution operators corresponding to purely
time-local kinetics \cite{l}, on opposite hand.

In our previous works~\cite{i1,i2,i3} the new
intermediate approach was suggested, so-called ``stochastic
representation of quantum interactions''. Its peculiarity is
that it involves time-local stochastic evolution equation
for density matrix of DS, with commutative (formally $c$-number
valued) random sources which generally undergo time-nonlocal
correlations. In contrary to phenomenological
``Langevin equations'' for DS variables,
this equation is exact, in the sense that quite definite and
transparent relations are established between complete
statistics of the random sources and private statistical
characteristics of the thermostat perturbed by classical
($c$-number valued) forces. These relations ensure adequate
construction of approximate models of the thermostat noise
in its connection with the thermostat induced dissipation.

In the present paper, most general formulation of this
approach is presented which unifies both quantum and
classical mechanics, at both Hamiltonian
and non-Hamiltonian interactions between DS and thermostat.
The results are illustrated by Brownian motion of classical
particle in classical gas, the ``trivial'' phenomemon which
still remains far from complete consideration (as well as
Brownian motion of gas molecules themselves, see \cite{i4}
and references therein).

\section{Stochastic representation of two-side interaction}

Let $\rho (t)\,$ be joint statistical operator
(distribution function, density matrix)
of combined system ``DS plus thermostat (environment)''
and $L(t)\,$ its evolution operator, so that
\begin{equation}
d\rho (t)/dt=L(t)\rho (t)                    \label{ee}
\end{equation}
We suppose that $L(t)\,$ has the bilinear form as follows:
\begin{equation}
L(t)=L_d(t)+L_b+L_{int}\,\,,\,\,\,
L_{int}=\sum_n \Lambda ^{d}_{n}\Lambda ^{b}_{n}\,,   \label{eo}
\end{equation}
where operators $L_d(t)$ and $L_b$ are
responsible for autonomous evolutions of DS and thermostat
(``bath''), respectively, while the pairs $\Lambda ^{d}_{n}$
and $\Lambda ^{b}_{n}$ for their interaction. The indexes
``d'' or ``b'' mark objects relating to DS or
thermostat, respectively.
Notice that all of $L_d$ and $\Lambda ^{d}_{n}$ commute
with all of $L_b$ and $\Lambda ^{b}_{n}$, since these two sets
of operators act onto variables of two different systems.

In general, $L_d(t)$ contains a part which describes
external observation of DS.
For instance, if DS undergoes quantum Hamiltonian evolution, then
\begin{equation}
L_d(t)\rho =\sum _k v_k(t)J_k\circ\rho
+i[\rho ,H_d(t)]/\hbar \,\,,                \label{qld}
\end{equation}
where $H_d(t)$ is Hamiltonian of DS; $\,J_k\,$ are some operators
belonging to DS and representing quantum variables under
observation; $\,v_k(t)$ are test functions related to $J_k\,$;
$\,[,]$ and $\,\circ \,$ mean commutator and Jordan symmetric
product, respectively:
\[
[A,B]\equiv AB-BA\,\,,\,\,A\circ B\equiv (AB+BA)/2
\]
In case of classical
Hamiltonian evolution, similarly,
\begin{equation}
L_d(t)\rho =\sum _k v_k(t)J_k\rho
+\{\rho ,H_d(t)\} \,\,,                \label{cld}
\end{equation}
with $\{,\}$ being the Poisson bracket, $\{A,B\}$ $\equiv $ $\frac
{\partial A}{\partial p}\frac {\partial B}{\partial q}-$ $\frac
{\partial A}{\partial q}\frac {\partial B}{\partial p}\,$ ($q\,$
and $p\,$ are canonical coordinates and momenta of DS), and $J_k$
some phase functions (i.e. functions of $q\,$ and $p\,$) under
observation. Then the trace, $\text{Tr\,}\rho \,$ (or, in
classical variant, integral over total phase space), differs from
unit and gives characteristic functional of the observables
$J_k(t)$, that is generating functional for their multi-time
correlators \cite{i1,i2,i3}.

Remembering that $L_d$ and $\Lambda ^{d}_{n}$ commute
with $L_b$ and $\Lambda ^{b}_{n}$,
it is not hard to justify the formal identity
\begin{equation}
\overleftarrow{\exp }\left\{\int\left[L_d(t)+L_b+
\sum \Lambda ^{d}_{n}\Lambda ^{b}_{n}
\right]dt \right\}=                     \label{id1}
\end{equation}
\[
=\overleftarrow{\exp }\left\{\int\left[
L_b+\sum \Lambda ^{b}_{n}\frac
{\delta}{\delta\alpha_n(t)}\right]dt \right\}
\]
\[
\times \,\overleftarrow{\exp }\left\{\int\left[
L_d(t)+\sum \alpha_n(t)\Lambda ^{d}_{n}
\right]dt \right\}\mid_{\alpha (t)=0}\,\,,
\]
where $\overleftarrow{\exp }\,$ denotes chronologically
ordered exponent. Because of the chronological ordering,
this formula is the mere consequence from the identity
$\exp(O_1O_2)=$ $\exp\left(O_1\frac {\partial}
{\partial \alpha}\right)$ $\exp(\alpha O_2)$
$\mid_{\alpha=0}\,$,
with $O_1$ and $O_2$ being two mutually
commuting operators ($[O_1,O_2]=0$).

From the other hand, if $\alpha _n(t) $ is a set of random
processes with a given characteristic functional
\[
\Psi\{f\}=\left\langle \exp\left[\int\sum
f_n(t)\alpha _n(t)dt\right]\right \rangle
\]
(the angle brackets mean statistical averaging),
then the average of arbitrary functional $\Phi\{\alpha\}$
can be formally represented as
\begin{equation}
\left\langle \Phi\{\alpha\}\right \rangle=
\Psi\left\{\frac {\delta}{\delta \alpha}
\right\}\Phi\{\alpha\}\mid_{\alpha=0}    \label{id2}
\end{equation}

Now, let us assume that more or less far in the past the joint
statistical operator $\rho (t)\,$ was factorized: $\,\rho (t_0)=$
$\rho _{d0}$$*\rho_{b0}\,$ (for instance, at $t_0=$ $-\infty $).
Then comparison between the Eqs. \ref{id1} and \ref{id2} clearly
results in two conclusions. First, the partial statistical
operator of DS, $\,\overline{\rho} _d(t)\equiv $
$\text{Tr\,}_b\,\rho (t)\,$, can be represented in the form
\begin{equation}
\overline{\rho }_d(t)\equiv \text{Tr\,}_b\,\rho (t)\,=
\left\langle \rho _d(t)\right\rangle\,\,,
                   \label{id}
\end{equation}
where $\,\rho _d(t)\,$ is a solution
(under initial condition $\rho _d(t_0)$ $=\rho _{d0}$)
to the stochastic evolution equation
\begin{equation}
d\rho _d(t)/dt=\left\{L_d(t)+\sum \alpha_n(t)
\Lambda ^{d}_{n}\right\}\rho _d(t)\,\,,  \label{se}
\end{equation}
with $\alpha_n(t)\,$ being random sources, and the angle brackets
$\,\left\langle \right\rangle \,$ standing for the average with
respect to these sources. Second, characteristic functional of
these sources is defined by
\begin{equation}
\left\langle\exp\left[\int\sum f_n(t)\alpha_n(t)dt\right]
\right\rangle=                            \label{cf}
\end{equation}
\[
=\text{Tr\,}_b\,\,\overleftarrow{\exp}\left\{\int
\left[ L_b+\sum f_n(t)\Lambda ^{b}_{n}
\right]dt\right\}\,\rho_{b0}
\]
Hence, from the point of view of DS its interaction
with thermostat is equivalent to its separate evolution
but disturbed by random sources and therefore described by
stochastic statistical operator of DS, $\,\rho _d(t)$.
The average $\left\langle \rho _d(t)\right\rangle\,$,
being a functional of the test functions $v_k(t)\,$
introduced in (\ref{qld}) and (\ref{cld}),
exactly reproduces not only current statistictical state of DS
but also all the multi-time correlators of the observables $J_k$.

The Eqs.\ref{id}-\ref{cf} give what we call stochastic
representation of interaction between two systems.
In this representation, the DS and thermostat evolutions
can be analysed separately. The cost of such possibility
is that both DS and thermostat should be considered under
arbitrary time-dependent perturbations ($\alpha_n(t)$
and $f_n(t)$, respectively). The examples~\cite{i1,i2,i3}
demonstrate that this cost may be not too large, since
when considering DS we become saved of many degrees of freedom
of thermostat and instead must deal with a few random
sources only. Importantly, like $c$-numbers, the sources always
commute one with another and with any other objects.
Besides, possibly, their exact statistics
which follows from (\ref{cf}), can be replaced by some
simplified (semi-phenomenological) statistical model.
Dependently on concrete contents of
operators $\Lambda ^{d}_{n}$ and $\Lambda ^{b}_{n}$,
the sources $\alpha_n(t)\,$ behave either as literally
classical random processes or as commutative ``ghost fields''
possessing unusual statistical properties
(see~\cite{i1,i2,i3} and below).

\section{Bilinear Hamiltonian interactions}
Let a quantum system ``DS plus thermostat'' has the Hamiltonian
\begin{equation}
H(t)=H_d(t)+H_b+H_{int}\,\,,\,\,\,
H_{int}=\sum_j B_j*D_j\,\,,         \label{blh}
\end{equation}
whose interaction part, $H_{int}\,$, is bilinear, with operators
$D_j$ and $B_j$ acting in different Gilbert spaces of DS and
thermostat, respectively, and therefore commuting one with
another. The corresponding joint evolution operator is $L(t)=$
$L_d(t)+$ $L_b+$ $L_{int}\,$, where $L_d(t)$ is given
by~(\ref{qld}),
 $L_b\rho =$ $i[\rho ,H_b]/\hbar\,$, and
\begin{equation}
L_{int}\rho =\sum \frac {i}{\hbar }[\rho ,D_j*B_j]  \label{lint}
\end{equation}
Because of $[D_j,B_j]=0$, for any pair $D_j$ and $B_j$ the
equalities take place:
\begin{equation}
[\rho ,D*B]=[\rho ,D]\circ B+D\circ [\rho ,B]    \label{eq}
\end{equation}
\[
=[\rho \circ B,D]+[\rho \circ D,B]
\]
Hence, the interaction part of the evolution operator has just
the bilinear structure as in~(\ref{eo}). We see that every
term $D_j$$*B_j\,$ from~(\ref{blh}) produces two terms
in $L_{int}\,$ in~(\ref{eo}), namely,
\begin{equation}
L_{int}=\sum _j\sum _{\sigma=1,2} \Lambda ^{d}_{j\sigma}
\Lambda ^{b}_{j\sigma}\,\,,                 \label{lint1}
\end{equation}
with the terms defined by
\begin{equation}
\Lambda ^{d}_{j1}O \equiv i[O,D_j]/\hbar \,,\,\,\,
\Lambda ^{b}_{j1}O \equiv B_j\circ O\,,\,\,  \label{lint2}
\end{equation}
\[
\Lambda ^{d}_{j2}O \equiv D_j\circ O\,,\,\,\,
\Lambda ^{b}_{j2}O \equiv i[O,B_j]/\hbar
\]
(here $O\,$ is an arbitrary operator).

Correspondingly, every term of $H_{int}\,$ results in
two random sources in~(\ref{se}), $\alpha _{j\sigma }(t)\,$
($\sigma =$ $1,2$). It is convenient to rename them
as $\alpha _{j1}(t)$ $=x_j(t)\,$, $\alpha _{j2}(t)$ $=y_j(t)$.
After that the Eq.\ref{se} takes the form
\begin{equation}
\frac {d\rho _d}{dt}=L_d(t)\rho _d+
\sum y_j(t)D_j\circ \rho _d+
\sum x_j(t)\frac {i}{\hbar}[\rho _d,D_j]  \label{se1}
\end{equation}
According to (\ref{cf}) and (\ref{lint2}), characteristic
functional of the sources in this equation looks as
\begin{equation}
\left\langle \exp \int
\sum [g_j(t)x_j(t)+f_j(t)y_j(t)]dt\right\rangle
=\text{Tr\,}_b\,\rho _b\,\,, \label{cf1}
\end{equation}
where operator $\rho _b\,$ (effective partial statistical operator
of thermostat) is defined as solution to the equation
\begin{equation}
\frac {d\rho _b}{dt}=L_b\rho _b+
\sum g_j(t)B_j\circ \rho _b+
\sum f_j(t)\frac {i}{\hbar}[\rho _b,B_j]\,,   \label{te1}
\end{equation}
with initial condition $\rho _b(t_0)$ $=\rho _{b0}$. This equation
describes separate evolution of thermostat, under its perturbation
by classical ($c$-number valued) forces $f_j(t)\,$ conjugated with
the observables $B_j$, and besides under measurement of the same
observables. The $\,g_j(t)\,$ are the probe (test) functions of
the measurement.

In the stochastic ``Liouville-Langevin equation'' (\ref{se1}), the
sources $x(t)$ play the role of usual (realistic) random forces
(potentials), while $y(t)$ are ``ghost'' ones. Indeed, if we put
on $g(t)$ $=0\,$ in (\ref{te1}) then the evolution described by
these equations becomes purely unitary. Therefore
$\text{Tr\,}_b\,\rho _b$ $=1$, and the Eq.\ref{cf1} yields
$\left\langle\exp\int f(t)y(t)dt\right\rangle $ $=1$.
Consequently, all the statistical moments of $y(t)\,$ and their
correlations between themselves are zeros. Nevertheless, the joint
statistical moments and cross correlations of $y(t)\,$ and
$x(t)\,$ are nonzero, being responsible, in particular, for
dissipation in DS \cite{i1,i2,i3}. According to (\ref{cf1}) and
(\ref{te1}),
\begin{equation}
\left\langle \prod _{j,m}x(t_j)y(\tau _m)\right\rangle =
\left [ \prod _m\frac {\delta }{\delta f(\tau _m)}
\left\langle \prod_j B(t_j,f)\right\rangle
\right ] _{f=0}\,,                 \label{cors0}
\end{equation}
where $B(t,f)$ are the thermostat observables $B$ considered as
functions of time and functionals of the Hamiltonian perturbation
characterized by the forces $f(t)\,$. For details and another
equivalent representations of the characteristic functional
see~\cite{i1,i2,i3}. It should be emphasized that due to the
causality principle any of the correlators (\ref{cors0}) turns
into zero if $\,\max _m\{\tau _m\}>$ $\max _j\{t_j\}$. In
addition, without loss of generality, the operators $B_j$ can be
defined in such a way that $\,\left\langle x_j(t)\right\rangle $
$=0\,$.

In case of thermodynamically equilibrium thermostat, the important
relations between pair correlators take place, namely,
\[
\left\langle x_j(\tau)x_m(0) \right\rangle = \int_0^\infty \cos
(\omega \tau )S_{jm}(\omega ) \frac {d\omega}{\pi}\,\,,
\]
\begin{equation}
\left\langle x_j(\tau)y_m(0) \right\rangle =\vartheta (\tau
)\times            \label{xyc}
\end{equation}
\[
\times \frac {2}{\hbar }\int_0^\infty \sin (\omega \tau )\tanh
\left( \frac {\hbar \omega}{2T} \right) S_{jm}(\omega ) \frac
{d\omega}{\pi}
\]
Here $T $ is the thermostat temperature, $\vartheta (\tau )$ is
Heavyside function, and $S_{jm}(\omega )$ is a non-negatively
defined spectral matrix. The connection between the two types of
correlators in (\ref{xyc}) merely expresses usual
fluctuation-dissipation theorem. In accordance with (\ref{xyc}),
if we neglected the sources $y(t)$, it would be equivalent to
infinite temperature.

\section{Non-Hamiltonian interactions}
We can obtain serious generalization of the representation
(\ref{lint1})-(\ref{lint2}) if replace (\ref{lint2}) by
\begin{equation}
\Lambda ^{d}_{j1}O \equiv i[O,D_j]/\hbar \,,\,\,\,
\Lambda ^{b}_{j1}O \equiv B^{\prime}_j\circ O
\,,\,\,           \label{lint3}
\end{equation}
\[
\Lambda ^{d}_{j2}O \equiv D^{\prime}_j\circ O\,,\,\,\,
\Lambda ^{b}_{j2}O \equiv i[O,B_j]/\hbar\,,
\]
thus involving not two but four different observables at any $j$,
with $D^{\prime}$ and $D$ belonging to DS
while $B^{\prime}$ and $B$ to thermostat.
Instead of (\ref{se1}) and (\ref{te1}), we come to the
representation as follows (summation over $j$ is taken in mind):
\begin{equation}
\frac {d\rho _d}{dt}=L_d(t)\rho _d+
y(t)D^{\prime }\circ \rho _d+
x(t)\frac {i}{\hbar}[\rho _d,D]\,,  \label{se2}
\end{equation}
\begin{equation}
\frac {d\rho _b}{dt}=L_b\rho _b+g(t)B^{\prime}\circ \rho _b+
f(t)\frac {i}{\hbar}[\rho _b,B]   \label{te2}
\end{equation}
This is accompanied by the same definition (\ref{cf1})
of the characteristic functional.

Obviously, the interaction defined by (\ref{lint1}) and
(\ref{lint3}) is not reducable to a bilinear Hamiltonian
interaction like (\ref{blh}), since any pair of the sources
($x(t)$ and $y(t)$) replaces not one but two different observables
of the thermostat ($B^{\prime}$ and $B$). As in (\ref{se2}),
$\,y(t)\,$ are ``ghost'' noise sources. The formula (\ref{cors0})
extends to this non-Hamiltonian case under substituting variables
$B^{\prime }(t,f)$ in place of variables $B(t,f)$ in
(\ref{cors0}). An exact contents of the correlators
$\left\langle\prod B^{\prime }(t_j,f)\right\rangle $ is directly
implied by the definition (\ref{cf1}).

The most general form of non-Hamiltonian interaction similar to
(\ref{se2}) and (\ref{te2}) is expressed by the formulas
\begin{equation}
L_{int}=\sum \{\Lambda ^d_j\Lambda ^{b\prime }_j+
\Lambda ^{d\prime }_j\Lambda ^{b}_j\}\,\,,   \label{lintg}
\end{equation}
\begin{equation}
\frac {d\rho _d}{dt}=L_d(t)\rho _d+
\sum \{y_j(t)\Lambda ^{d\prime }_j+
x_j(t)\Lambda ^d_j\}\rho _d\,,            \label{seg}
\end{equation}
\begin{equation}
\frac {d\rho _b}{dt}=L_b\rho _b+
\sum \{g_j(t)\Lambda ^{b\prime }_j+
f_j(t)\Lambda ^b_j\}\rho _b\,,        \label{teg}
\end{equation}
again accompanied by (\ref{cf1}), where $\Lambda ^d_j$ and
$\Lambda ^b_j$ are some generators of unitary (phase volume
preserving) individual evolutions of DS and thermostat,
respectively, while $\Lambda ^{d\prime }_j$ and $\Lambda ^{b\prime
}_j$ generate non-unitary phase volume exchange between DS and
thermostat (their mutual observation one for another).

\section{Brownian motion}
Examples of the bilinear Hamiltonian interaction can be
found in \cite{i1,i2,i3}. As concrete example of the previous
non-Hamiltonian case, consider Brownian motion
of classical particle in classical gas.

\subsection{Brownian particle in a gas}
Let $\bm{R}$, $\bm{V}$, $\bm{P}$ and $M$ be position vector,
velocity, momentum and mass, respectively, of Brownian particle
(BP), $\,\bm{r}_j$, $\bm{v}_j$ , $\bm{p}_j$ and $m\,$ be analogous
quantities of gas particles, and the latters interact with BP by
mean of potentials $U(\bm{r}_j-\bm{R})$. The joint evolution
operator is merely the Liouville operator of the system ``DS and
thermostat (gas)'':
\begin{equation}
L=L_d(t)+L_g+ \sum \bm{F}(\bm{r}_j-\bm{R})\cdot \left( \frac
{\partial}{\partial \bm{P}} -\frac {\partial}{\partial
\bm{p}_j}\right )\,,                     \label{bpi}
\end{equation}
where $\bm{F}(\bm{r})$ $\equiv $ $-\nabla U(\bm{r})$, and $L_g$ is
the Liouville operator of the gas itself. Generally $L_d$ includes
three terms:
\begin{equation}
L_d(t)=i\bm{k}(t)\cdot \bm{V}-
\bm{V}\cdot\frac {\partial}{\partial \bm{R}}-\bm{F}_{ext}(t)
\cdot \frac {\partial}{\partial \bm{P}}\,\,,   \label{nld}
\end{equation}
with $\bm{F}_{ext}(t)$ being an external force applied to BP, and
$\bm{k}(t)\,$ the test (probe) function for observing BP's
velocity ($i\bm{k}(t)\,$ plays the same role as $v(t)$'s in
(\ref{cld})).

Let us reduce expression (\ref{bpi}) to the form of
non-Hamiltonian bilinear interaction (\ref{eo}), (\ref{lintg}).
First, make the non-canonic change of variables, namely, consider
positions of gas particles in terms of the relative distancies
$\bm{r}_j-\bm{R}$ and redenote the latters as $\bm{r}_j$. Under
this change, $L_g\,$ preserves its form, but
$\bm{F}(\bm{r}_j-\bm{R})$ transforms into $\bm{F}(\bm{r}_j)$, and
$\frac {\partial}{\partial \bm{R}}$ transforms into $\frac
{\partial}{\partial \bm{R}}$ $-\sum \frac {\partial}{\partial
\bm{r}_j}$. Second, introduce the operator
\begin{equation}
L_b=L_g-\sum \bm{F}(\bm{r}_j)\cdot
\frac {\partial}{\partial \bm{p}_j}   \label{nlb}
\end{equation}
Formally, $L_b$ describes the gas in presence of immovable
scatterer fixed at zero point. After that (\ref{bpi}) transforms
into
\begin{equation}
L=L_d(t)+L_b+L_{int}\,\,,    \label{bpi1}
\end{equation}
\begin{equation}
L_{int}=\sum \bm{F}(\bm{r}_j)\cdot \frac {\partial}{\partial
\bm{P}}+\bm{V}\cdot \sum \frac {\partial}{\partial
\bm{r}_j}                    \label{nint}
\end{equation}

Hence, $L_{int}$ has just the structure (\ref{lintg}),
and we can identify the $\Lambda $'s (replacing their indexes
by natural vectorial notations):
\[
\bm{\Lambda }^d=-\frac {\partial}{\partial \bm{P}}\,\,,\,\,\,
\bm{\Lambda }^{b\prime }=-\sum \bm{F}(\bm{r}_j)\,,
\]
\begin{equation}
\bm{\Lambda }^{d\prime }=\bm{V}=\frac {\bm{P}}{M}\,\,,\,\,\,\,
\,\bm{\Lambda }^b=\sum \frac
{\partial}{\partial \bm{r}_j}    \label{intop}
\end{equation}
Correspondingly, the Eqs. \ref{seg} and \ref{teg} look as
\begin{equation}
\frac {d\rho _d}{dt}=\left\{L_d(t)+
\bm{y}(t)\cdot \bm{V}-\bm{x}(t)\cdot \frac {\partial}{\partial
\bm{P}}\right \}\rho _d\,,         \label{sebp}
\end{equation}
\begin{equation}
\frac {d\rho _b}{dt}=\left\{L_b-
\bm{g}(t)\cdot \sum \bm{F}(\bm{r}_j)+
\bm{f}(t)\cdot \sum\frac {\partial}{\partial \bm{r}_j}
\right\}\rho _b\,,       \label{tebp}
\end{equation}
to be accompanied by (\ref{nld}) and (\ref{nlb}). For the
characteristic functional (\ref{cf1}) let us introduce the
designation
\begin{equation}
\Xi \{\bm{g},\bm{f}\}\equiv
\left\langle e^{ \int [\bm{g}(t)\cdot \bm{x}(t)
+\bm{f}(t)\cdot \bm{y}(t)]dt }\right\rangle
=\text{Tr\,}_b\,\rho _b         \label{cfbp}
\end{equation}
Clearly, in (\ref{sebp}) $\bm{x}(t)$ is actual random Langevin
force produced by the gas. In (\ref{tebp}), factual physical
dimensionality of the ``force'' $\bm{f}(t)$ is velocity.

Eventually we are interested in the characteristic functional of
the velocity and displacement of BP, that is
\begin{equation}
\overline {\Theta }\{i\bm{k},\bm{F}_{ext}\} \equiv \left\langle
e^{\int i\bm{k}(t)\cdot \bm{V}(t)dt }\right\rangle  \label{vcf}
\end{equation}
It can be found in two steps:
\begin{equation}
\Theta \{i\bm{k},\bm{F}_{ext},x,y\}\equiv \int \rho _d\,dRdV\,\,,
\label{cfbp1}
\end{equation}
\begin{equation}
\overline {\Theta }\{i\bm{k},\bm{F}_{ext}\} = \left\langle \Theta
\{i\bm{k},\bm{F}_{ext},x,y\}\right\rangle  \label{cfbp2}
\end{equation}
The first step is very easy, due to that the coefficients of the
stochastic Liouville operator on right-hand side of Eq.\ref{sebp}
(first-order differential operator) do no depend on $R$ and are
linear functions of $V$. Supposing, without loss of generality,
that initially BP was fixed at zero point at its phase space,
$\,\bm{R}(t_0)$ $=0\,$, $\,\bm{V}(t_0)$ $=0\,$, i.e. $\rho _{d0}=$
$\delta(\bm{R})\delta(\bm{V})\,$, we can obtain
\begin{equation}
\Theta \{i\bm{k},\bm{F}_{ext},x,y\}=    \label{cfbp3}
\end{equation}
\[
\exp \left\{\int_{t_2>t_1}
\left [i\bm{k}(t_2)+\bm{y}(t_2)\right]\cdot
\left [ \bm{F}_{ext}(t_1)+\bm{x}(t_1)
\right]\frac {dt_1dt_2}{M} \right\}
\]
In opposite, the second step, that is averaging (\ref{cfbp3}) with
respect to random trajectories $x(t)$ and $y(t)$, is rather
nontrivial problem.

\subsection{Path integral formulation}
Of course, we can rearrange $\Phi $ and $\Psi $ in the identity
(\ref{id2}) and besides replace $\alpha _n$ by $c_n\alpha_n$ with
$c\,$ being any nonzero coefficients. Therefore let us write
\begin{equation}
\overline {\Theta }\{i\bm{k},\bm{F}_{ext}\} = \left[\Theta
\left\{i\bm{k},\bm{F}_{ext},\frac {\delta }{\delta ig},\frac
{\delta }{\delta f} \right\}\Xi \{ig,f\}\right]_{g=f=0}
              \label{pi1}
\end{equation}
Due to simple specific structure (\ref{cfbp3}) of the functional
$\Theta $, we can easy transform the latter differential
expression into integral one if use functional analogue of the
formal identities
\begin{equation}
\exp\left(i\tau\frac {\,\partial ^2}{\partial a\partial
b}\right)\delta(a)\delta(b)=
\frac{1}{2\pi\tau}\exp\left(\frac{iab}{\tau}\right)\,, \label{pi2}
\end{equation}
\begin{equation}
\left[\exp\left(i\tau\frac {\,\partial ^2}{\partial a\partial
b}\right)\Phi(a,b)\right]_{a=b=0}=             \label{pi3}
\end{equation}
\[
=\frac{1}{2\pi\tau}\int da\int db\,
\exp\left(\frac{iab}{\tau}\right)\Phi (a,b)
\]
The final result of subsequent natural algebraic manipulations is
the path integral
\begin{equation}
\overline {\Theta }\{i\bm{k},\bm{F}_{ext}\} = \int DgDf\,\,
\Xi\{ig,f\}\times           \label{pi4}
\end{equation}
\[
\exp\left\{i\int\left[M\bm{f}(t)\frac{dg(t)}{dt}+
\bm{k}(t)\bm{f}(t)+\bm{g}(t)\bm{F}_{ext}(t)\right]dt\right\}\,,
\]
with $\,DgDf\equiv $ $\prod_t[Mdg(t)df(t)/2\pi]\,$ being path
differential in some reasonable space of $g(t)$,$f(t)$
trajectories. For brevity, here and below we omit the central dot
symbolizing scalar product.

\subsection{Limit of ideal gas}
In the simplest case, let the gas be ideal, i.e. there are no
interactions between its atoms and besides no initial statistical
correlations between them:
\[
L_g=-\sum _{k=1}^N \bm{v}_j\frac {\partial}{\partial
\bm{r}_j}\,\,, \,\,\,\rho _{b0}=\prod _{k=1}^N \frac
{W_0(\bm{r}_k,\bm{v}_k)}{\Omega }\,,
\]
with $\Omega $ and $N$ being total volume and number of atoms of
gas, respectively. Naturally, the initial one-particle
distribution function can be supposed homogeneous
(space-independent at least far from BP), for instance,
$W_0(\bm{r},\bm{v})$ $=W_0(\bm{v})$, while $W_0(\bm{v})$ arbitrary
even (spherically symmetric) velocity distribution. In particular,
the reasonable choice for $W_0(\bm{v})$ is usual Maxwellian
distribution corresponding to thermally equilibrium gas with a
definite temperature.

In the ``thermodynamic limit'' $N\rightarrow \infty $,
$\Omega\rightarrow \infty $, $N/\Omega \rightarrow \nu \,$ ($\nu
=$const), statistics of the thermostat noise gets the general
Poissonian form (i.e. becomes infinitely divisible):
\begin{equation}
\Xi \{\bm{g},\bm{f}\}=\exp \left(\nu\, Q\{\bm{g},\bm{f}\}\right)
     \label{ops}
\end{equation}
Here functional $Q\{\bm{g},\bm{f}\}$ characterizes contribution to
the overall noise from one atom:
\[
Q\{\bm{g},\bm{f}\}\equiv\int [W(t,\bm{r},\bm{u})-
W_0(\bm{r},\bm{u})]\,d\bm{r}d\bm{u}\,\,,
\]
with function $W(t,\bm{r},\bm{v})\,$ undergoing the equation
\begin{equation}
\frac {dW}{dt}=\left\{-\bm{g}(t)\bm{F}(\bm{r})
+[\bm{f}(t)-\bm{v}]\frac {\partial}{\partial \bm{r}}
-\frac {\bm{F}(\bm{r})}{m}\frac {\partial}{\partial \bm{v}}
\right\}W                  \label{ope}
\end{equation}
and the initial condition $W(t_0,\bm{r},\bm{v})$
$=W_0(\bm{r},\bm{v})$. Hence, analysis of Eqs. \ref{tebp} and
\ref{cfbp}, in company with Eq.\ref{nlb}, reduces to the
one-particle problem.

In Eq.\ref{ope} the first right-hand term says that an atom is
observed by BP. The next two terms say that it is dynamically
perturbed by BP, as if an atom would have the Hamiltonian $H=$
$[\bm{p}-m\bm{f}(t)]^2/2m$ $+U(\bm{r})$. Hence, the ``force''
$\bm{f}(t)$ representing velocity of BP acts like an effective
time-varying vector potential.

Alternatively, we can consider this one-particle problem in terms
of the factual velocity of an atom, $\,\bm{u}$ $\equiv $
$\bm{p}/m-\bm{f}(t)\,$, instead of its momentum $\bm{p}$ or
$\bm{v}=$ $\bm{p}/m$. Then Eq.\ref{ope} transforms into
\begin{equation}
\frac {dW}{dt}=\left\{-\bm{g}(t)\bm{F}(\bm{r})
-\bm{u}\frac {\partial}{\partial \bm{r}}
+\left[\frac {df(t)}{dt} -\frac {\bm{F}(\bm{r})}{m}\right]
\frac {\partial}{\partial \bm{u}}\right\}W          \label{ope1}
\end{equation}
In agreement with the above mentioned condition $\bm{V}(t_0)$
$=0\,$, we can assume that $\,\bm{f}(t_0)$ $=0\,$ and that initial
condition for the $W\,$ treated as a function of the factual
velocity $\bm{u}=$ $\bm{p}/m-\bm{f}(t)\,$ is
$W(t_0,\bm{r},\bm{u})$ $=W_0(\bm{r},\bm{u})$. Because integration
over $\bm{u}$ in (\ref{ops}) is indifferent to shift of the
integration variable, $\bm{u}\rightarrow $ $\bm{u}+\bm{f}(t)$, the
function $W(t,\bm{r},\bm{u})$ in the characteristic functional
(\ref{ops}) can be mentioned as solution to Eq.\ref{ope1}.

\section{Discussion and resume}
The above example once again demonstrates (see also
\cite{i1,i2,i3}) how the stochastic representation formalism can
help in reducing many-particle problems to one- or few-particle
problems. It will be interesting to find out how far this approach
conducts in analyzing factual statistics of Brownian motion beyond
conventional Gaussian approximation.

Of special interest is one-dimensional motion which arises in the
problem about massive piston sliding in (infinitely long) tube
filled with an ideal gas \cite{chls}. It was rigorously proved
(see \cite{chls} for the review and original results) that in the
limit of infinitely heavy BP (piston) its velocity tends
expectedly to the Ornstein-Uhlenbeck random process possessing
Gaussian statistics.

From the other hand, our (mathematically non-rigorous) attempt
\cite{i4} (see also \cite{i5}) to reconsider derivation of
Boltzmann kinetics of classical weakly non-ideal gas, in the
Boltzmann-Grad limit, demonstrated that statistics of Brownian
motion (self-diffusion) of gas atoms is essentially non-Gaussian
because of low-frequency fluctuations in diffusivities of the
atoms (uncertainty of the diffusivities).

Hence, the task naturally arises about non-Gaussian corrections to
statistics of piston's motion at large but finite ratio $M/m\,$.
In principle, it is sufficient to assume symmetrical situation
when initially both ``left-hand'' and ``right-hand'' gases are in
identical (mirror symmetrical) states. In order to keep constant
piston's velocity relaxation time, the gas density $\nu\,$ (number
of atoms per unit tube length) should be maintained proportionally
to $M/m\,$.

Due to the symmetry, in Eq.\ref{ops},
\[
 Q\{g,f\}= Q_{+}\{g,f\}+Q_{+}\{-g,-f\}\,\,,
\]
where $Q_{+}\{g,f\}$ is connected to solution of Eq.\ref{ope} or
Eq.\ref{ope1} for the right-hand gas atom by the same
Eq.\ref{ops}. The source of possibly non-trivial corrections (as
well as non-trivial mathematical difficulties, see \cite{chls}) is
the recollisions between piston and atoms.

Indeed, in an infinitely long container (tube), let before a
collision occuring at $t=0$ the piston's velocity, $V(t)\,$, and a
``right-hand'' atom's velocity, $v(t)\,$, were equal to $V_0\,$
and $v_0\,$, respectively, of course, satisfying $v_0<V_0\,$.
After collision the atom's velocity equals to $v=$ $2V_0-v_0$,
while the distance, $\delta R(t)$, between piston and atom changes
as
\[
\delta R(t)=vt-R(t)\,\,,\,\, R(t)=\int_0^tV(t^{\prime})dt^{\prime}
\]

The recollision will happen if $\delta R(t)=0$ at $t>0$. Suppose
that after the collision the atom runs on the left, i.e. $v=$
$2V_0-v_0$ $<0\,$, which means that it was overtaking the piston,
i.e. $V_0<0\,$. Besides let $V(t)$ is symmetric Ornstein-Uhlenbeck
random process. Then, obviously, it is quite impossible that
recollision will not happen, since this would mean that symmetric
Brownian path $R(t)$ for the whole time $t>0$ remains confined
below the straight line $vt\,$ with negative slope $v<0$. In
opposite, the recollision is inevitable. Consequently, its
probability is not less than the probability that $2V_0$ $<v_0<$
$V_0<0$, that is (at nearly equilibrium situation) it has at least
the order on $\sqrt{m/M} $.

Such estimates prompt that probabilities of multiple recollisions
also are substantial. Due to all the recollisions, relatively slow
atoms are accelerated by piston when it in its turn is accelerated
by dominating relatively fast atoms. In other words, slow atoms
``dress the piston in a coat'' thus randomly changing its
effective mass. Seemingly, this must imply some fluctuations in
relaxation time and diffusivity of the piston. From the point of
view of \cite{i4,i5} this effect associates with that space-drift
velocity of probability of some collisions chain is the
center-of-mass velocity of particles taking part in this chain.
Whether low-frequency fluctuations in diffusivity of Brownian
particle wandering in one-dimensional ideal gas really exist is
the question for separate consideration.


\begin{acknowledgments}
I am grateful to Dr. I.Krasnyuk for useful conversations.
\end{acknowledgments}

\begin{verbatim}
\end{verbatim}




\end{document}